\documentclass[12pt]{article}
\pdfoutput=1
\usepackage{geometry} 
\usepackage{graphicx}
\usepackage{amssymb,amsmath}
\usepackage{jneurosci}
\usepackage{tabularx}

\title{Network structure determines patterns of network reorganization during adult neurogenesis}
\author{Casey M. Schneider-Mizell\textsuperscript{1}, Jack M. Parent\textsuperscript{2}, Eshel Ben-Jacob\textsuperscript{6},\\ Michal Zochowski\textsuperscript{*1,3,4,5}, Leonard M. Sander\textsuperscript{*1,5}\\
\\
Departments of \textsuperscript{1}Physics, \textsuperscript{2}Neurology, \textsuperscript{3}Biophysics Research Division,\\ \textsuperscript{4}Neuroscience Graduate Program, and\\ \textsuperscript{5}Michigan Center for Theoretical Physics,\\ University of Michigan, Ann Arbor, Michigan 48109, USA;\\ \textsuperscript{6}School of Physics and Astronomy, Tel-Aviv University, Tel Aviv, Israel}

\begin{document}
\maketitle

\begin{abstract}
	New cells are generated throughout life and integrate into the hippocampus via the process of adult neurogenesis. Epileptogenic brain injury induces many structural changes in the hippocampus, including the death of interneurons and altered connectivity patterns. The pathological neurogenic niche is associated with aberrant neurogenesis, though the role of the network-level changes in development of epilepsy is not well understood. In this paper, we use computational simulations to investigate the effect of network environment on structural and functional outcomes of neurogenesis. We find that small-world networks with external stimulus are able to be augmented by activity-seeking neurons in a manner that enhances activity at the stimulated sites without altering the network as a whole. However, when inhibition is decreased or connectivity patterns are changed, new cells are both less responsive to stimulus and the new cells are more likely to drive the network into bursting dynamics. Our results suggest that network-level changes caused by epileptogenic injury can create an environment where neurogenic reorganization can induce or intensify epileptic dynamics and abnormal integration of new cells.

\end{abstract}

\section{Introduction}

Continuous introduction of new neurons via adult hippocampal neurogenesis is thought to assist memory formation and storage \cite{Aimone:2006p9548,Silva:2009p9622,Zhao:2008p8217}. While many details of neurogenesis at the molecular and cellular level are known \cite{Balu:2009p7199}, the effect of network environment on the integration of newly born cells remains unresolved. In this paper, we use a computational model to investigate how the established network structure can alter patterns of neural integration and how these changes affect the evolution of structure and dynamics. 

Hippocampal neuroblasts arise from progenitor cells in the subgranular zone of the dentate gyrus (DG) and migrate a short distance into the granule layer \cite{Altman:1965p10072,Kaplan:1977p10079,Cameron:1993p10076,Kuhn:1996p10080}. They develop synaptic inputs from GABAergic interneurons whose activity promotes the development of neural processes \cite{OverstreetWadiche:2006p7561,Ge:2007p8202}. After weeks, they form glutamergic synapses onto mossy cells and interneurons \cite{Toni:2008p6481}. Studies have found new granule cells to be preferentially activated by stimulation \cite{RamirezAmaya:2006p6345} and to have sensitive synaptic plasticity \cite{SchmidtHieber:2004p6108}. These findings show neurogenesis to be promoted by and respond specifically to activity in the established network.

Epileptogenic injury can significantly alter the neurogenic niche. Mossy fiber sprouting increases the number of granule cell synaptic outputs, hilar cell death reduces inhibition, and granule cell dispersion changes the spatial structure of the granule cell layer \cite{Buckmaster:1997p8766}. Additionally, seizures form a different pattern of activity. The rate of neurogenesis after status epilepticus has been observed to increase \cite{Parent:1997p10081}. Some neuroblasts migrate ectopically into the hilus and, unlike in the healthy brain \cite{Parent:1997p10081,Scharfman:2000p10085,Parent:2006p8150}, appear to receive synapses from other granule cells \cite{Pierce:2005p10082}. 

Previous computational work on neurogenesis has focused on learning and memory \cite{Aimone:2009p7085,Becker:2005p7369,Chambers:2004p7370,Deisseroth:2004p7371,Wiskott:2006p7372}. Here, we consider the general question of how the network activity and structure affects new cell integration and long term network activity. We build a computational model with simplified neurogenesis rules and study spatial patterns of integration as a function of structural characteristics of the network environment. We consider two levels of inhibition: 1) when the ratio of inhibitory to excitatory cells represents that observed in the healthy hippocampus, and 2) that of epileptic hippocampus with a reduced inhibitory population. Under normal conditions, we find an optimal network topology which is dynamically and structurally robust to the addition of new neurons. For other network structures or decreased inhibition, neurogenesis leads to significant structural and dynamical changes in the underlying network. We thus postulate that there is a range of normal conditions in which neurogenesis could enhance network performance, but if the underlying network structure is pathological, it can \emph{worsen} the pathology. 
	
\section{Materials and Methods}
	
The simulation is broken into three primary aspects: creation of the established network structure, the neuronal dynamics, and the integration of new neurons. See Table~\ref{parameterTable} for a consolidated list of all parameters.
\subsection{Initial network formation}
We create the initial networks prior to neurogenesis by placing 1000 excitatory neurons and either 200 (normal conditions) or 100 (reduced inhibition) inhibitory neurons at random on a two-dimensional square lattice with 40 lattice sites per side and periodic boundaries. A lattice site can hold at most one excitatory cell. Inhibitory cells are placed on a location independent of its occupancy status. We specify that each excitatory cell will be connected to, on average, 3.5\% of the other excitatory cells. 

\begin{figure}[htbp]
\begin{center}
\scalebox{0.5}{\includegraphics{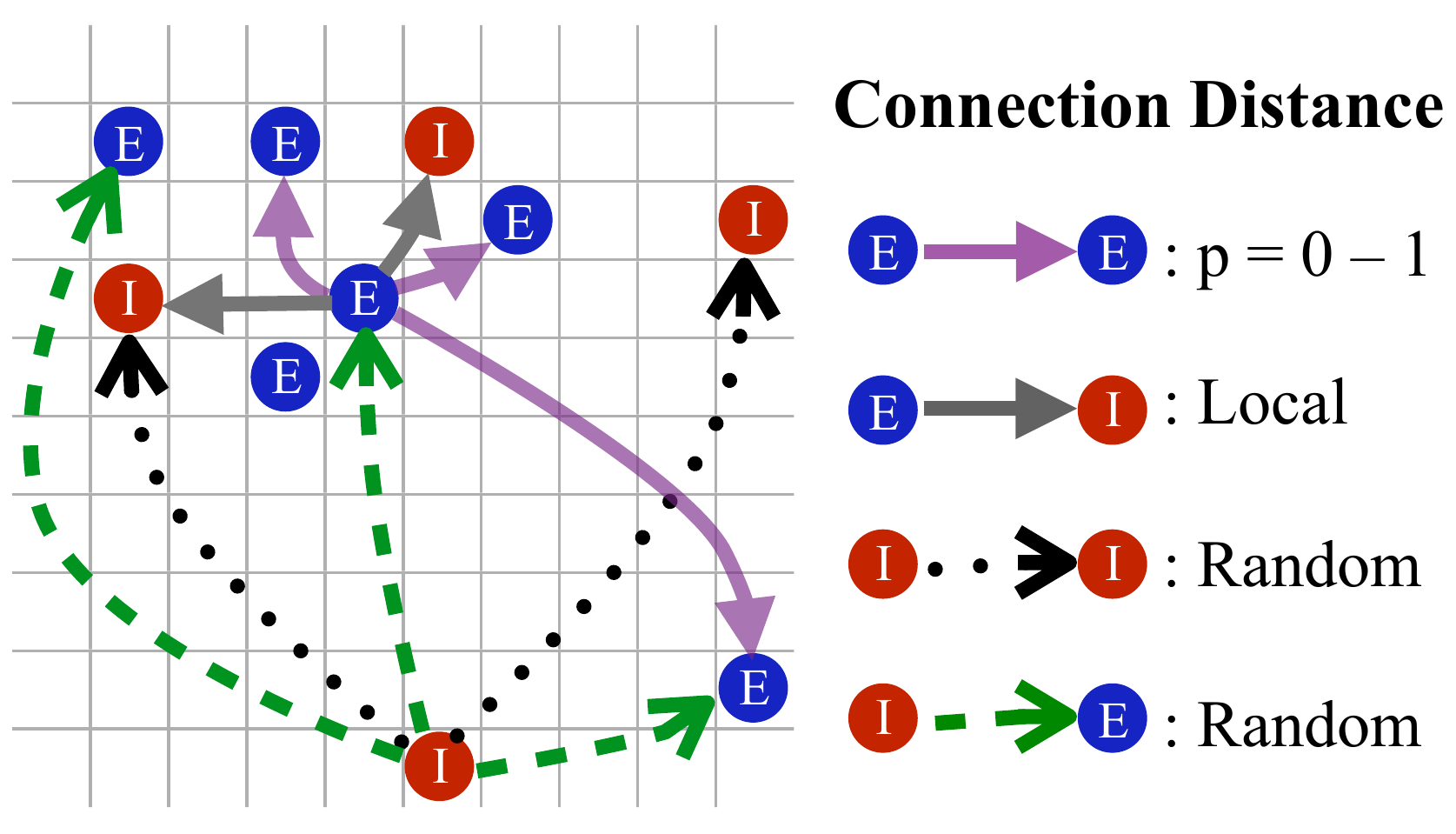}}
\caption{A cartoon of the connection scheme for the established network. Connections from inhibitory cells to either excitatory or other inhibitory cells are random and excitatory cells connect only to nearby inhibitory cells. For excitatory-excitatory connections, the ratio of local to random connections is determined by a rewiring probability $p$, which we vary.}
\label{netdiagram}
\end{center}
\end{figure}

We consider random networks with variable rewiring to represent connections between neurons. This type of model is well known to possess a small-world regime \cite{Watts:1998p2416}. See Figure~\ref{netdiagram} for a cartoon of the following process. For each cell we determine its downstream (axonal) connections by looking within a radius (R) containing twice the desired average number of out-connections and connecting to each with probability 1/2. For each connection, we rewire it to a randomly chosen excitatory cell anywhere in the network with a probability, $p$, which we vary. For $p=0$, all connections are local, and for $p=1$ all connections are random.  Each excitatory neuron is also connected to, on average, four inhibitory neurons through the same procedure as described above, without random rewiring. All excitatory-to-inhibitory connections are therefore local. Inhibitory cells send out-connections to, on average, four other inhibitory cells and 100 excitatory cells. These targets are selected purely at random. We adopted this connectivity structure to roughly represent that of the CA3 layer of the hippocampus \cite{Jablonski:2007p6470}. However, the reported results are robust to changes in these parameters (see Supplementary Data).

\subsection{Network dynamics}
We represent network activity with integrate-and-fire dynamics with stochastic spontaneous firing. While integrate-and-fire is only a very rough approximation of the dynamics of real neurons, it is sufficient for the purposes of our model, for which connectivity, not detailed dynamics, is paramount.

In the integrate-and-fire scheme, the $i$th cell has a voltage $V_i$ which follows:
$$
\frac{dV_k}{dt} = (I_{e/i} + E_k - \alpha_k V_k) + \sum_j w_{jk} A_{jk} I_{syn}^j(t+\tau_{delay}),
$$
where $I_{e/i}$ is the global excitability of excitatory/inhibitory neurons, depending on what class of neuron $k$ belongs to. The initial potentials are distributed uniformly at random between 0 and 1. The cellular and network parameters are adopted from \cite{Jablonski:2007p6470}. We use $I_i = 0.6$ and $I_e = 0.73$. The external stimulus is denoted by $E_k$, which is $0$ for unstimulated cells and 0.4 for those which receive external stimulus. Stimulus occurs at the five centermost excitatory cells. The membrane leak constant is $\alpha_k$, drawn for each cell from a uniform distribution between 1 and 1.3. The network adjacency matrix elements are denoted by $A_{jk}$, which is 1 if neuron $k$ sends an output to neuron $j$ and 0 otherwise.  The synaptic weight is denoted by $w_{jk}$. Synaptic weight is based only on the class of the neurons involved, with $w_{ee} = 0.2$, $w_{ei} = 0.4$, $w_{ie} = -0.4$, and $w_{ii} = -0.7$. The negative sign represents inhibition. Synaptic weight and excitability parameters are inspired by \cite{Jablonski:2007p6470}, where they were tuned to give controlled dynamics. Finally, the synaptic current from the $j$th neuron is $I_{syn}^j(t)$, where $I_{syn}(t)$ comes from a double exponential based on the time $t$ since neuron $j$ fired plus a signaling delay $\tau_{delay}=0.08$: 
$$
I_{syn}(t) = e^{-t/\tau_S} -e^{ - t/\tau_F }
$$
where $\tau_S = 3 ms$ and $\tau_F = 0.3 ms$. This spike form is taken from \cite{Netoff:2004p9177}.

The voltage dynamics is numerically solved using the Euler method. Membrane potentials are capped below by 0 and if $V_j > 1$, the neuron $j$ fires. All cells also have a probability of 0.0003 per ms of spontaneously firing, providing a mean background firing rate per cell of 0.3Hz. When a neuron fires, its potential is reset and held at 0 for a refractory time of 8 ms.

\subsection{Neurogenesis and network reorganization	}
In order to represent neurogenesis, we introduce a new neuron every 350 ms after an initial 1,000 ms of simulation time. A new cell is placed on a randomly chosen unoccupied lattice site. We then proceed to form inward (dendritic) and outward (axonic) connections. We form inward connections by compiling a list of all inward connections of neighboring cells (within radius $R/2$) to the newly added one. This list denotes the possible in-connections. Inputs are drawn from this list by assigning a score $g_i$ to each neuron on the list based on firing rate $f_i$ and a small amount of random jitter $\omega$ drawn uniformly between 0 and 1:
$$
g_i =  a \frac{f_i}{ \max_{\{i\}}{f_i}} + (1-a) \omega.
$$
The parameter $a$ determines the amount of randomness in the selection process. We use $a = 0.8$ to wash out small differences in firing activity and preserve large ones. The neurons with the highest values of $g$ are selected to be inputs to the new cell, until the appropriate number of inputs is reached. Our results are not sensitive to modest changes in $a$ (see Supplementary Data).

Similarly, the output connections are drawn from downstream targets of nearby cells. Connections are made to a random subset of possible outputs until the new cell has the same average number of connections as the original network. 

During the initial stages of its dynamics the network can undergo rapid rewiring. At intervals of 200 ms of simulation time, all output connections which do not result in a sufficient number of coincident firing events (defined as a post-synaptic cell firing within 10 ms after the pre-synaptic cell fires) are broken and new connections are again chosen at random. After 2000 ms of simulation time, the new cell either dies if its firing rate is less than 100 times the spontaneous background firing rate (0.3Hz) or matures and ceases to undergo changes in its connections. When a new cell survives, the total population is kept constant by killing a randomly chosen mature cell.

Birth and maturation rates are chosen so that there is little interaction between two immature cells and the network activity is well sampled. Spatial activity distributions were not observed to change considerably for the parameters used here, except in response to new cells. Network activity can thus be well measured by 2000 ms of activity, even though it is not a physiologically relevant time period.

Except where noted, simulations are performed by first running dynamics on the stimulated initial network for 1000 ms to establish baseline activity. We then add 500 new cells as described above, and stop the simulation 1000 ms after the last new cell has had an opportunity to mature. Each data point represents 50 realizations. All simulations, data analysis, and plotting were performed in Matlab 7.7.0 (Mathworks).

\subsection{Metrics of network activity}
\subsubsection{Spatial measurements}

To measure and visualize the firing activity, survivorship, and number of reconnection events as a function of location, we break the space into a $20\times20$ grid of squares covering the network space, so that each square represents four lattice sites. Activity and reconnection events are both measured as the average value among all cells that fall within a given square. For survivorship, we report the average number of cells that survive within a given square. 

\subsubsection{Radial component}

To measure the mean radial component of output connections, we first create a vector for each new cell that represents the average direction of all outputs. This is defined as $$\vec{v_i} = \frac{1}{k_i} \sum_j A_{ji} \frac{\vec{x_j}-\vec{x_i}}{ |{\vec{x_j}-\vec{x_i}} |}$$ where  $k_i = \sum_j A_{ji}$ is the number of out-connections from neuron $i$. We then measure the inward radial component by taking the dot product of this mean direction with the inward-pointing unit radial vector $\hat{r}$  at the location of the $ith$ neuron and normalizing by the magnitude of $\vec{v_i}$:  $\hat{r} \cdot  \vec{v_i} /{| \vec{v_i} |}$. This is averaged for all surviving new neurons in each simulation. Values close to one indicate highly radial directionality among new neurons' outputs. Values near zero are consistent with random directions.

\subsubsection{Synchronous bursting}

We use an existing measure of synchronicity of bursting based on interspike time differences \cite{Tiesinga:2004p9532}. We first make an ordered list of all spike times $t_\nu$ for all excitatory cells. The measure B is defined as
$$
B = \frac{1}{\sqrt{N}} \left( \frac{ \sqrt{\langle \tau_\nu^2 \rangle - \langle \tau_\nu \rangle^2} }{ \langle \tau_\nu \rangle } - 1 \right)
$$
where $\tau_\nu = t_\nu - t_{\nu+1}$ is the time difference between subsequent firing events for on the excitatory network and angle brackets indicate averages over all such time differences. For large $N$ and all neurons firing as independent Poisson processes, $B=0$, and for synchronous bursting $B=1$.

\subsubsection{Spike order}

The relative intraburst order of new and old cells is measured as the time difference between the onset of new cells bursting and old cells bursting. We define the onset of a burst as the time at which the number of new or old cells firing simultaneously increases past four. We make a list of the onset of all new and old bursts. For each burst of new cells, we record the time difference to the closest old burst onset. The convention is chosen such that an old burst that leads a new burst has a negative time difference.

\subsection{Pilocarpine-induced status epilepticus}
Adult male Sprague-Dawley rats were pretreated with atropine methylbromide (5 mg/kg intraperitoneally [IP]; Sigma), and 15-minutes later were given pilocarpine hydrochloride (340 mg/kg IP; Sigma) to induce SE. Seizures were terminated with diazepam (10 mg/kg) after 90 minutes of SE as previously described \cite{Parent:2006p7576}. Controls received the same treatments as experimental animals except that they were given saline in place of pilocarpine.

\subsection{Retrovirus production and injection}
Replication-incompetent recombinant RV vectors were pseudotyped by co-transfection of GP2-293 packaging cell line (Clontech,) with plasmids containing the RV vector (RV-CAG-GFP-WPRE, gift of S. Jessberger and F. Gage) and vesicular stomatitus virus (VSV)-G envelope protein (Clontech). The supernatant containing RV was harvested and filtered through a 0.45-$\mu$m pore size filter (Gelman Sciences) and centrifuged in a Sorvall model RC 5C PLUS at 50,000xg at 4$^\circ$C for 90 minutes. The RV-containing pellet was resuspended in 1X PBS, aliquoted, and subsequently stored at -80$^\circ$C until use. The concentrated RV titer was determined using NIH 3T3 cells and found to be approximately 1-5 x 108 CFU/mL. For intrahippocampal RV injections, animals were anesthetized with a ketamine/xylazine mixture and placed on a water-circulating heating blanket. After positioning in a Kopf stereotaxic frame, a midline scalp incision was made, the scalp reflected by hemostats to expose the skull, and bilateral burr holes drilled. RV vector (2.5 $\mu$L of viral stock solution was injected into the left and right dentate gyri over 20 minutes each using a 5 $\mu$L Hamilton Syringe, and the micropipette left in place for an additional 2 minutes. Coordinates for injections (in mm from Bregma and mm depth below the skull) were caudal 3.9; lateral 2.3, depth 4.2.

\subsection{Tissue processing, immunohistochemistry, and microscopy}
 Four weeks after SE, animals were deeply anesthetized and perfused with 4\% paraformaldehyde (PFA). The brains were removed, postfixed for 4-6 hours in 4\% PFA, cryoprotected in 30\% sucrose and frozen. Coronal sections (40 $\mu$m thick) were cut with a freezing mictrotome and fluorescence immunohistochemistry was performed on free-floating sections \cite{Parent:1997p10081,Parent:1999p10456} using rabbit anti-GFP primary antibody (1:1000, Invitrogen) and Alexa 488-conjugated anti-rabbit IgG secondary antibody (1:400, Invitrogen). Images were captured using a Zeiss LSM 510 confocal microscope.

\section{Results}
The purpose of the simulations was to understand how the network connectivity structure may influence spatial patterns of network augmentation during the adult neurogenesis and the dynamical patterns of the resulting network. We investigated the spatial patterns of neurogenesis, their established connectivity patterns and their dynamics with relation to the existing cells as a function of the network topology. This study was carried out for two cases: 1) when the inhibition resembles that observed during the normal conditions and 2) when the inhibition is reduced to represent the situation after loss of hilar interneurons as in temporal lobe epilepsy.

\subsection{Properties of the initial network structure}
\begin{figure}[htbp]
\begin{center}
\scalebox{0.5}{\includegraphics{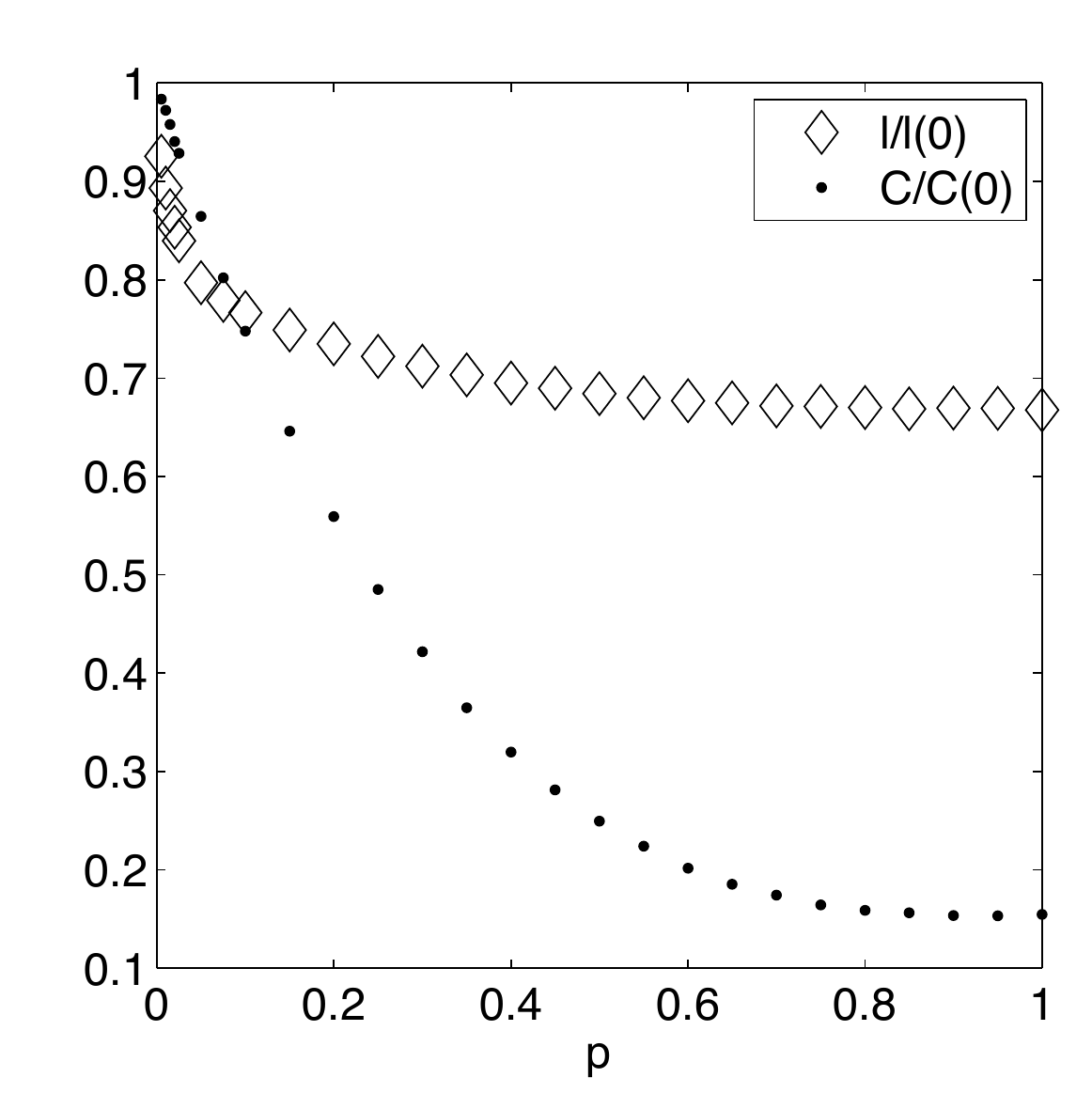}}
\caption{Average mean path length (diamonds) and clustering coefficient (dots) as a function of rewiring probability $p$, normalized by the value for $p=0$. All data points are averages over 20 simulated networks. The path length drops sharply as $p$ increases, but the clustering coefficient has much slower decrease. The small-world regime is defined by having large clustering but small path length, which is found for rewiring probabilities near $p=0.1$.}
\label{LCPlot}
\end{center}
\end{figure}

The topology of the initial network in our model is inspired by Watts-Strogatz networks \cite{Watts:1998p2416}. The network structure depends on a parameter, $p$, which controls the rewiring probability of a given connection between excitatory cells. This parameter allows the connectivity of excitatory-excitatory networks to range from purely local ($p=0$) to purely random ($p=1$), while preserving many other aspects of the network such as the degree (i.e. cell connectivity) distribution. Between the local and global extremes is a small-world topology. Small-world networks are characterized by having short average path lengths between any two nodes, but a high probability that the neighbors of a given node are connected to one another, as measured by the clustering coefficient \cite{Newman:2003p90}. One can determine the range of values of $p$ for which the network exhibits small-world characteristics by measuring the clustering coefficient and mean path length (Figure~\ref{LCPlot}). We observe small-world structure in our excitatory networks around $p=0.05$--0.2.

\subsection{Activity and augmentation patterns of the network in response to external stimulus}
\begin{figure}[htbp]
\begin{center}
\scalebox{0.6}{\includegraphics{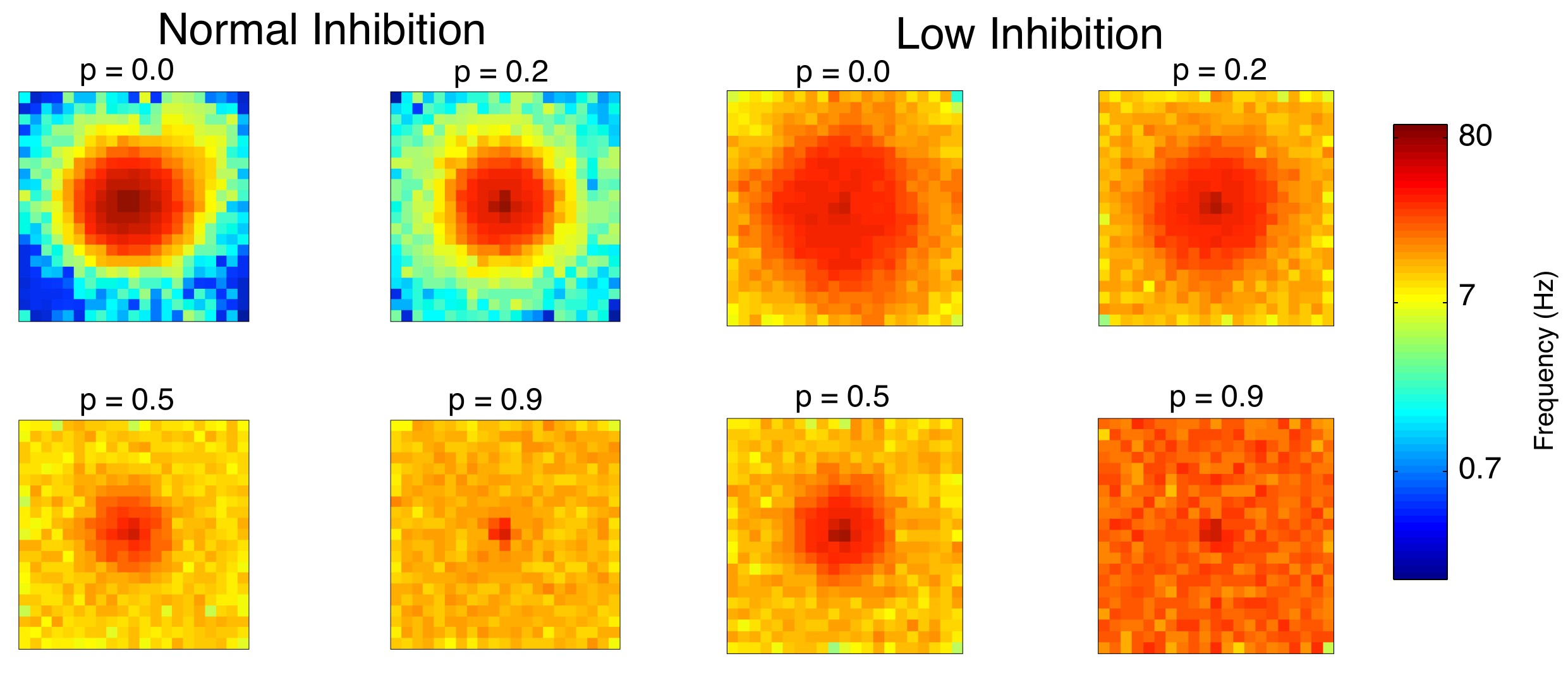}}
\caption{Spatial distribution of mean firing rate after introducing 500 cells for normal (left) and low inhibition (right). Note log scale. Shown are four example rewiring probabilities for each level of inhibition. Observe that the effective localization of activity for normal inhibition and low $p$ that is not present for other parameter values.}
\label{actMap}
\end{center}
\end{figure}

	We first consider how the network activity patterns respond to external stimulus for different network topologies. We stimulated five excitatory neurons positioned in the center of the network by applying an additional constant current. The spatial distribution of activity in the network after the addition of new cells as a function of rewiring probability $p$ and the amount of inhibition in the original network is shown in Figure~\ref{actMap}. For established networks with $p<0.2$ and normal inhibition, we see highly localized activation patterns (note log color scale), while global connectivity patterns and/or diminished inhibition produce global patterns of activation. 
	
\begin{figure}[htbp]
\begin{center}
\scalebox{0.6}{\includegraphics{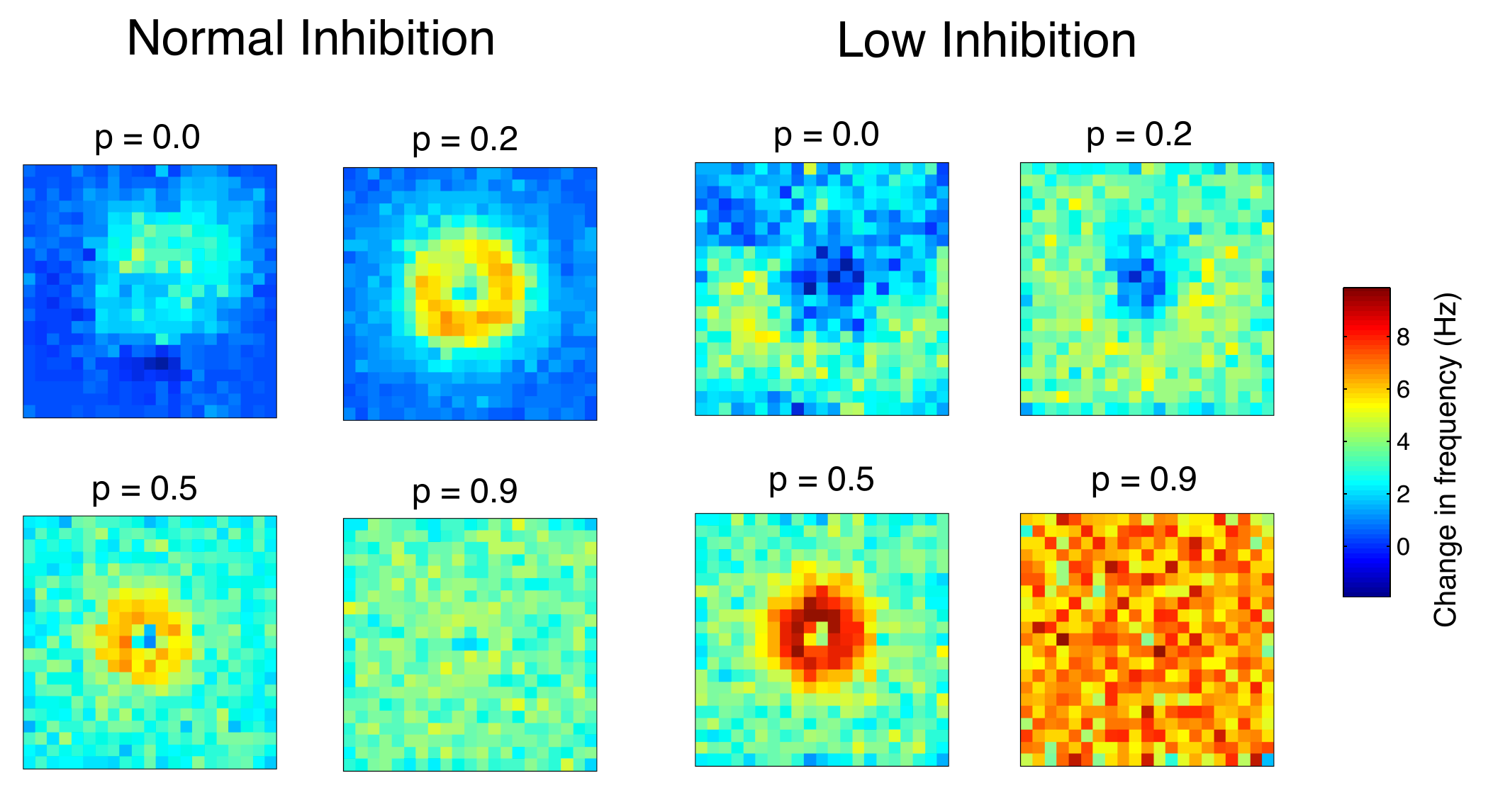}}
\caption{Spatial distribution of the change in firing rate after introducing 500 cells for normal (left) and low inhibition (right). For normal inhibition and low $p$, the network activity changes only near the stimulated region. However, for larger $p$, thus more random connections, or decrease in inhibition, activity increases spread through the network. For low inhibition and low rewiring, some regions show little change because the stimulus is already driving some cells near their maximum rate.}
\label{deltaF}
\end{center}
\end{figure}

	If we examine only the changes in firing rate (Figure~\ref{deltaF}), we see that new cells are causing an increase in the firing rate that remains near the stimulated region for normal inhibition and low $p$. This confirms that our neurogenesis process can sharpen the response of our network to a stimulus. Increasing the rewiring and decreasing inhibition each result in a global increase of firing rate, with amounts varying depending on the details of the network. 
		
\begin{figure}[htbp]
\begin{center}
\scalebox{0.65}{\includegraphics{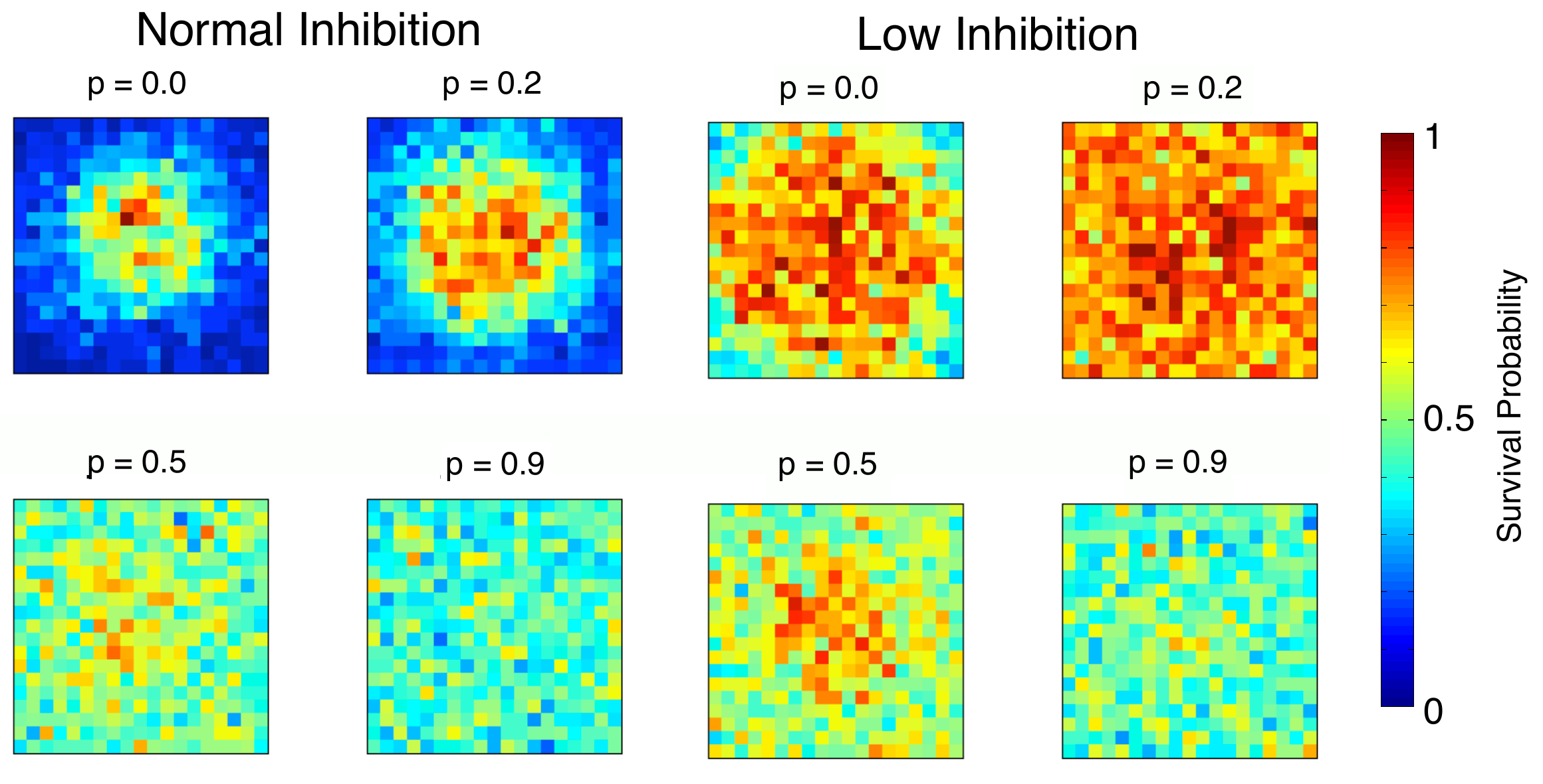}}
\caption{Spatial distribution of the survival probability from fifty simulations for normal (left) and low inhibition (right). Note that survival is highly localized for the same values that gave localized activity and increased firing rate.}
\label{locMap}
\end{center}
\end{figure}

	We next consider the spatial patterns of survivorship of cells that we added. We observed that, as above, for local connectivity and normal inhibition, the location of active network reorganization (i.e. high survival probability for new cells) is focused around the stimulation area (Figure~\ref{locMap}). This effect diminishes for more global connectivity and decreased inhibition. This indicates that to maintain specific activity after incorporation of new cells, as we expect in normal hippocampal conditions, the established network must have relatively local connectivity of the original network and sufficiently high levels of inhibition.

\begin{figure}[htbp]
\begin{center}
\scalebox{0.5}{\includegraphics{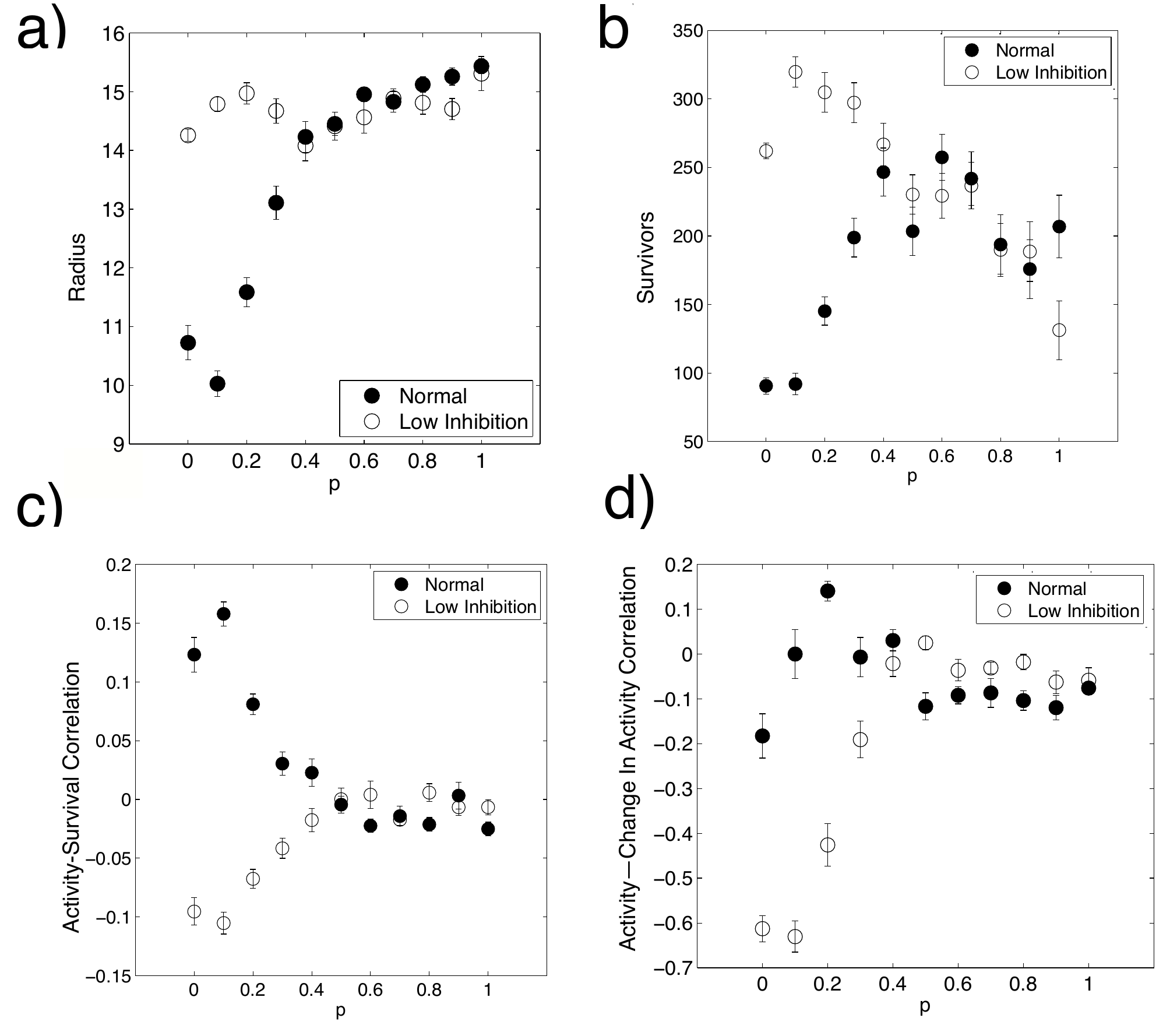}}
\caption{a) Mean radius of survivors as a function of $p$ and inhibition, measured in lattice sites. The width of the network is 40 sites, so a distance near 15 is consistent with uniform distribution. Again, normal inhibition and small-world rewiring cause neurogenesis to respond with more focus. b) Average number of survivors out of 500 introduced. Low values of rewiring and normal inhibition give low rates of incorporation. c) Spatial correlation between activity at the beginning of simulation and survival probability. While under normal inhibition new cells stay in the stimulated areas, for low inhibition cells can survive in distant locations because of the greater changes in activity. d) Spatial correlation between initial activity and the change in activity after all cells have been introduced. Again, the peak value occurs for normal inhibition and a rewiring probability that puts the network into the small-world regime.}
\label{spatialStats}
\end{center}
\end{figure}

	We can develop a general picture of what occurs for each network structure by measuring the mean radius of survivors in the network calculated from the center of the stimulation and mean number of new cells that survive (Figure~\ref{spatialStats}a and b). To see how initial activity affects survivorship and changes in activity, we also plot the space-averaged correlation between initial network activity and survival probability as well as the correlation between initial activity and change in activity (Figure~\ref{spatialStats}c and d). These measurements are done for both high inhibition regime and low inhibition. Note that the correlation between activity and change in activity is artificially low, since the locations that have the highest activity may be unable to fire faster due to the refractory time setting a maximum frequency.
		
	We observe that under normal inhibition the smallest number of added cells survives when the network is in the small-world regime. Furthermore, the highest correlation between initial activity and survival probability also occurs for the same regime. Peak correlation for change in activity is at a slightly different location than survivorship, but still in the range where the established network is small-world. Increasing the rewiring parameter $p$, and thus the number of random connections, and decreasing the inhibition have similar effects of reducing correlation and increasing both the number of surviving cells and the spatial extent of where they survive and activate the network.
	
	We also considered the effect of an increase in the number of connections between excitatory neurons (see Supplementary Data). We find that increasing the excitatory to excitatory connectivity from 3.5\% to 4.5\% while retaining normal inhibition does not qualitatively change the pattern of rewiring above $p=0.2$. However, for $p \leq 0.1$, new cells enter the network with spatial patterns similar to that observed under reduced inhibition. There appears to be a fast crossover between low-inhibition-like behavior and normal-inhibition-like behavior. However, because integration patterns mimic low and high inhibition cases in other respects, we focus on those two aspects of the network here.
	
\subsection{Innervation patterns of incorporated neurons}
The newly born neurons are thought to have higher structural plasticity, enabling them to rapidly form and abolish synaptic connections to other cells. This potentially allows the cells to optimally integrate themselves into the existing networks. In our simulation this reconnection occurs when an immature neuron is unable to generate a sufficient number of downstream coincident firing events. We observed that the number of activity dependent reconnection events per surviving cell is highest for global network topologies (Figure~\ref{orientationPlot}d). Furthermore, networks with lowered inhibition exhibit lower rates of reconnection events for $p<0.5$ due to the overall activity being higher in the network with lowered inhibition. New cells therefore do not seek specific targets, indicating a lower potential controllability of the process, as well as a possible reason for significant modifications of the hippocampal network structure during epileptogenesis as shown in Figure~\ref{orientationPlot}b.

\begin{figure}[htbp]
\begin{center}
\scalebox{0.8}{\includegraphics{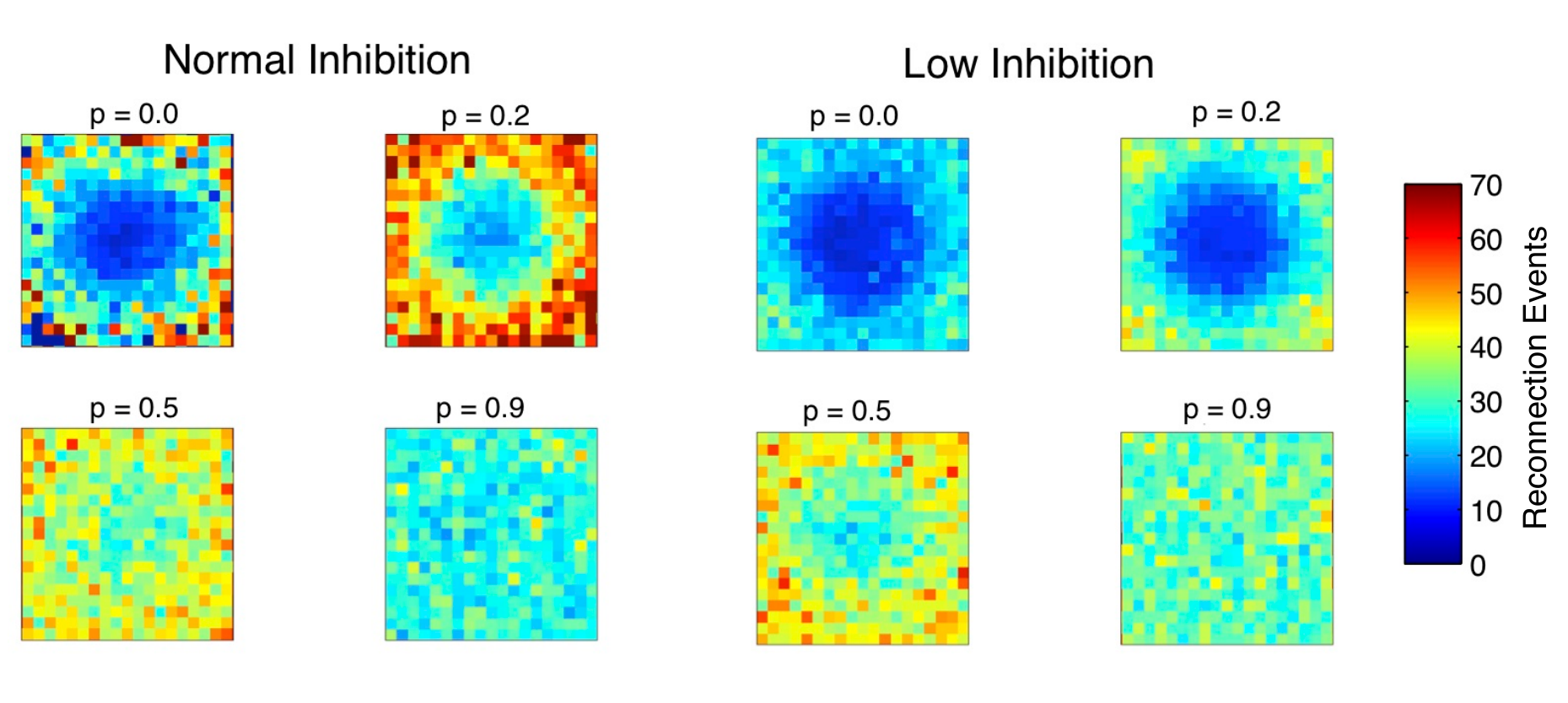}}
\caption{Spatial distribution of mean reconnection events per surviving cell for four example rewiring probabilities. Note the presence of an ``easy" place near the stimulus to incorporate into the network at low $p$. Large numbers of reconnection events mean that a new cell is not able to quickly find neighbors that it can functionally innervate.}
\label{reconMap}
\end{center}
\end{figure}

We have also looked at the spatial distribution of reconnection events as a function of $p$ and inhibition (Figure~\ref{reconMap}). Under normal inhibition and for relatively local networks, cells that lie outside the immediate stimulated area undergo significantly more reconnection events. Established cells within the stimulated region are more likely than remote cells to respond to firing events from new cells, making it an easier region to wire into.  This effect becomes less dramatic and eventually goes away for more random network topologies as well as for networks with diminished inhibition.

Another important feature of neurogenesis is the innervation pattern of the new cells. We consider each synaptic connection to be spanned by a long axon and short dendrites, such that the direction of a connection is roughly that of the innervating axon. For low values of $p$, a clear pattern of radially oriented outputs toward the stimulation site can be observed among introduced cells (Figure~\ref{orientationPlot}c). This disappears for large $p$. To quantify this, we plotted the average radial component for the output directions of all surviving new cells (Figure~\ref{orientationPlot}d). The networks having normal inhibition levels that develop the most highly ordered (i.e. directed toward stimulus) connectivity are those having small-world properties. More random networks continue developing without any strongly ordered directionality. Additionally, low inhibition networks are not able to develop the same degree of directionality as normal networks.

\subsection{Firing dynamics}
\begin{figure}[htbp]
\begin{center}
\scalebox{0.5}{\includegraphics{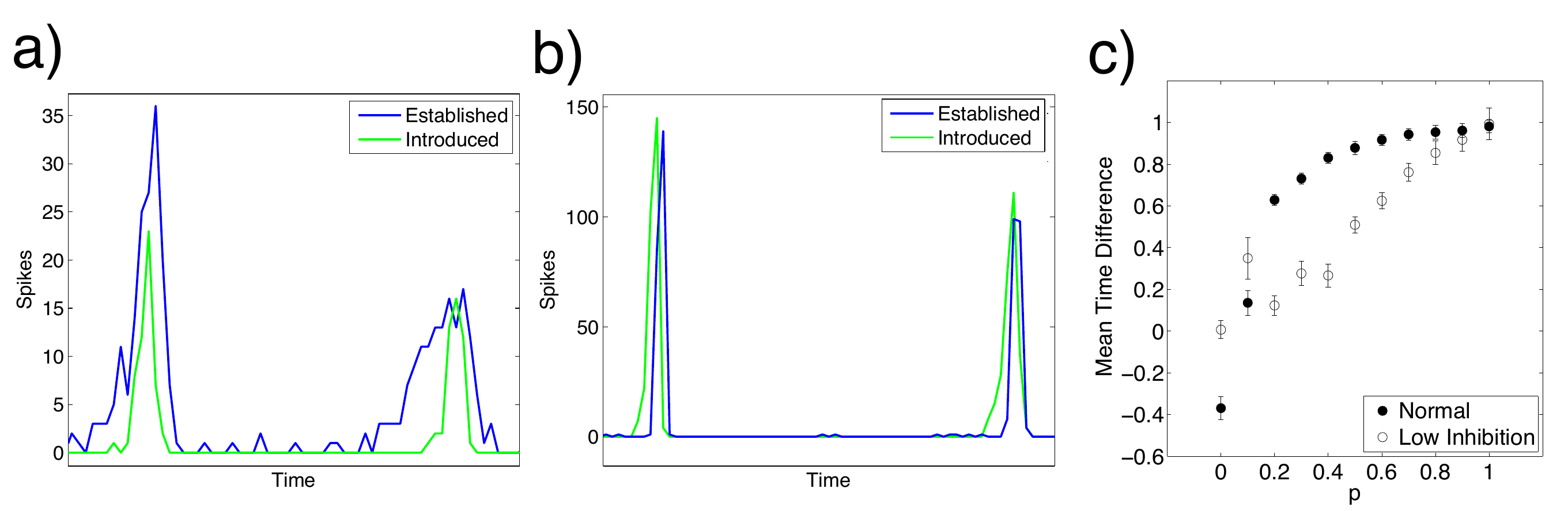}}
\caption{Comparison of age-dependent burst times. a) An example of firing activity for low rewiring ($p=0$), showing new neurons firing (green) within bursts of established neurons (blue). b) An example of firing activity for high rewiring ($p=0.7$). Note that the new neurons (green) spike first, followed by the established network (blue). c) Mean time difference between onset of bursts among new neurons and established neurons. A positive value implies that new neurons fire first. The time difference is measured from a time window at the end of the simulations. For low rewiring, new neurons are part of bursts, but for high rewiring they drive bursts.}
\label{dynDiff}
\end{center}
\end{figure}

We also investigated differences in the firing patterns of new and established cells in reorganized networks. If we look at traces of the cumulative activity patterns, we can see that for small-world networks, activation of new neurons overlaps with that of the already existing cells (Figure~\ref{dynDiff}a) without a significant lead-lag pattern emerging between the activations of the two populations. For random networks, bursts from new cells tend to lead those of the original ones (Figure~\ref{dynDiff}b). 
	
To quantify this observation, we measured the time difference between the onset of bursts of the old and new subpopulations (Figure~\ref{dynDiff}c). This time difference is negative when old cells lead and positive when new cells lead. We restrict ourselves to data taken from the end of the simulation. We see that, in most cases, new cells will lead bursts in the established network. The only case where the established network consistently leads bursts is for low rewiring and normal inhibition. The transition between which cell population leads occurs when the network is in the small-world regime.

Notably, for low inhibition there is never a regime with a negative time difference. However, the new cell lead time is generally smaller for low inhibition because the established network requires less activity to be driven.

\subsection{Robustness of network activity}
Finally we wanted to investigate how robust is the network dynamics to adding new cells. To address this, we ran simulations that terminated when 500 new cells survived and were incorporated into the network, instead of 500 being introduced to the network. Because of this, some simulations were run for longer time period than others. We quantified the activity patterns with the mean frequency of the network and a measure of synchronous bursting, $B$.

Local networks are much more robust to activity-dependent reorganization than those with more random connections (Figure \ref{robustPlot}). In all cases, established networks with low $p$ tolerate the incorporation of activity-seeking neurons with only modest increases in activity and synchronous firing. Networks with more global connections start with slightly less activity, but are not able to accommodate the same number of neurons without having increasing firing rate and spontaneous bursting. Decreasing inhibition does not qualitatively change this behavior, although the mean firing rate is generally higher.

\section{Discussion}
In this work, we investigated how network topology can mediate changes in network structural and dynamical reorganization during neurogenesis. To that effect we constructed a simplified model incorporating activity dependent augmentation rules and investigated outcomes as a function of topology.  

We observed that the networks with sufficiently high inhibition and small-world topology are robust to incorporation of the new cells in terms of structural and dynamical properties. In these networks pre-existing spatial and temporal firing patterns are reinforced by new neurons, with survival and connectivity patterns highly stimulus dependent. This is consistent with the idea that hippocampal neurogenesis contributes to memory formation via high stimulus dependent plasticity. Networks with less inhibition or more long range connections are unstable to activity-dependent augmentation. Activity becomes more globally synchronized and new cells incorporate more randomly and in stimulus independent fashion. Taken together, these results indicate that structural network differences alone can be sufficient to cause qualitatively different outcomes after neurogenic reorganization. Moreover, the changes that occur in the hippocampal network after the onset of epilepsy are similar to the changes in our network model that induce global synchrony and erratic incorporation of new neurons.

Considerable experimental \cite{FarioliVecchioli:2008p7598,Cao:2004p9748,vanPraag:1999p9750} and computational work \cite{Aimone:2009p7085,Becker:2005p7369,Chambers:2004p7370,Deisseroth:2004p7371,Wiskott:2006p7372} has focused on the role of neurogenesis in learning. While results are not entirely consistent in their details, there is a general consensus that the highly plastic new cells increase the ability of the hippocampus to store, maintain, and retrieve spatial memories. In order to do this, new cells have to be able to respond with specificity to stimulus \cite{Silva:2009p9622}.

We found our network to have stable, focused reorganization by neurogenesis for a small range of rewiring probabilities near $p=0.1$ and normal levels of inhibition. For networks having that topology, the correlation of network activity with survival location of the new cells is maximized, the  processes of new cells are most ordered, and the network retains overall dynamical properties after adding new cells. Interestingly, our analysis of the initial network structure for this parameter range corresponds to the high-clustering, low-path-length small-world regime. It is commonly thought that many parts of the brain are wired as small-world networks, and quantitative estimates of the DG connectivity imply that it is actually a small-world network \cite{DyhrfjeldJohnsen:2007p6934}.

Numerous changes have been observed in the DG of epileptic subjects. Two of the most important structural changes are hilar cell death, in which interneurons and mossy cells die, and mossy fiber sprouting, in which granule cells develop increased axonal arbors and new synapses. Detailed computational modeling of the DG has observed that these structural changes in the hippocampus can make the DG more likely to exhibit global synchrony \cite{DyhrfjeldJohnsen:2007p6934,Morgan:2008p7862}.

Experimentally observed neurogenic outcomes differ greatly. Diminished specificity of response of the post-neurogenesis epileptic network to stimuli in the form of directionality and location is observed in a variety of experiments. Widening of the granule cell layer and ectopic integration of neuroblasts into the dentate hilus are examples of wide dispersal in epilepsy \cite{Parent:2006p7576}. While cells migrate abnormally due to changes in motility cues \cite{Gong:2007p8141}, how they alter the network dynamics is still an open question. The orientation of these new cells is also highly random, in considerable change from the ordered picture seen in healthy granule cells. Additionally, neurogenesis accelerates after seizures in terms of both increased proliferation \cite{Bengzon:1997p10068,Parent:1997p10081} and faster functional integration \cite{OverstreetWadiche:2006p8453}. Experimental observations have also found that experimental epilepsy reduces learning ability \cite{Stafstrom:1993p9913,Holmes:1997p9981,Jessberger:2007p8299}.

\begin{figure}[htbp]
\begin{center}
\scalebox{0.6}{\includegraphics{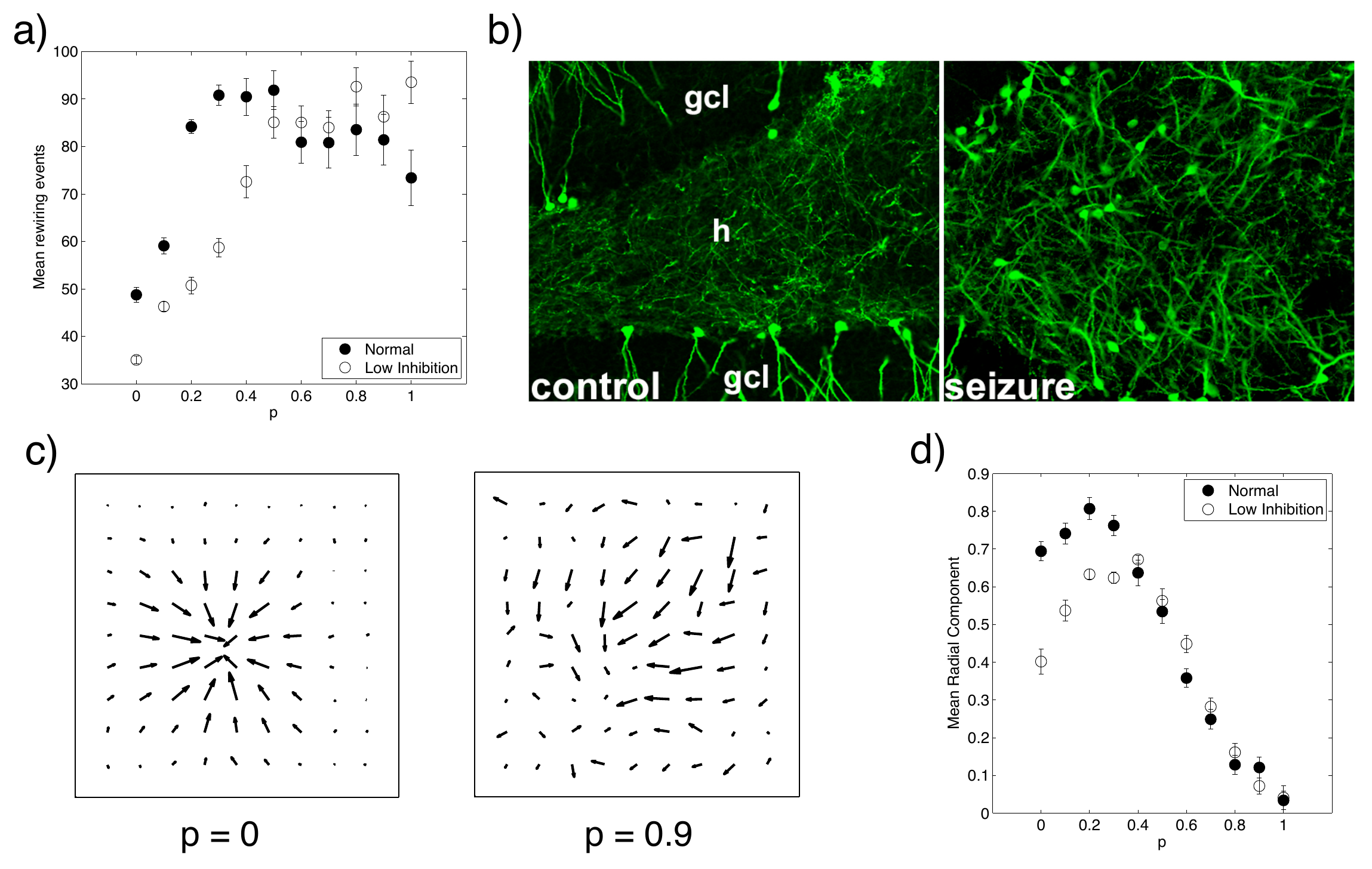}}
\caption{Connection patterns of surviving cells. a) Average number of rewiring events. Networks with low inhibition require less searching to effectively integrate into the network. b) New granule cell (green) integration into the hippocampus is orderly and with consistent directions in normal conditions. The hilus is denoted by ``h" and the granule cell layer by ``gcl". After seizures are induced, granule cells can integrate and orient in a more random manner. See Methods for experiment description. c) Mean direction of output connections as a function of location for two different values of rewiring. Both have normal inhibition. For low rewiring, the connections are aligned radially, whereas for higher rewiring the orientation becomes more random. b) Average radial component of output connections. High values mean that connections are forming toward the stimulated region as in the $p=0$ example, whereas values near zero are consistent with random, as in the $p=0.9$ example.}
\label{orientationPlot}
\end{center}
\end{figure}

We have modeled a portion of the structural network changes found in human and experimental temporal lobe epilepsy by implementing reduced numbers of inhibitory cells and changes in the network topology. In our model, networks which deviate from the small-world and  normal inhibition regime, undergo significant structural and dynamical modifications during augmentation and tend to develop towards higher activity levels, global synchrony and a disordered response to the stimulus. 
Increasing rewiring induces much greater shifts away from stimulus specificity, with the additional characteristics of activity spreading widely through the network and activity-survival correlation becoming zero. The result is similar to the randomly oriented post-seizure neurons shown in Figure~\ref{orientationPlot}b. Decreasing inhibition as well causes even low values of rewiring to result in widespread activity and survival probability, as well as the increased survival rates observed experimentally. 
	
Experiments on rats found that suppressing neurogenesis after status epilepticus reduced the duration and frequency of occurrence of recurrent seizures \cite{Jung:2004p9227,Jung:2006p10095}, consistent with the picture that neurogenesis can reinforce and spread epileptic dynamics. In both cases the probability of developing seizures after SE was also reduced. Taken together, it suggests that some injuries that did not induce seizures created an environment that was pushed toward pathological dynamics by neurogenesis. In our model, increases in rewiring probability make the network less stable to reorganization by neurogenesis, but decreasing inhibition can directly change dynamics.

Finally, our results predict that under pathological (epileptic) conditions the newly incorporated cells can exhibit shifts in their activity patterns compared to neurons in the established network. We did not observe these differences under normal conditions. This is possibly related to the observation that hubs (nodes with many more connections than average) can induce epileptic dynamics \cite{Morgan:2008p7862}. The activity-seeking nature of new cells is a mechanism that could create these hub cells.  The relative dynamics of newly added neurons into pathological and normal networks would be an interesting avenue for experimental study to gain insight into the possible epileptogenic nature of neurogenesis.

\begin{figure}[htbp]
\begin{center}
\scalebox{0.6}{\includegraphics{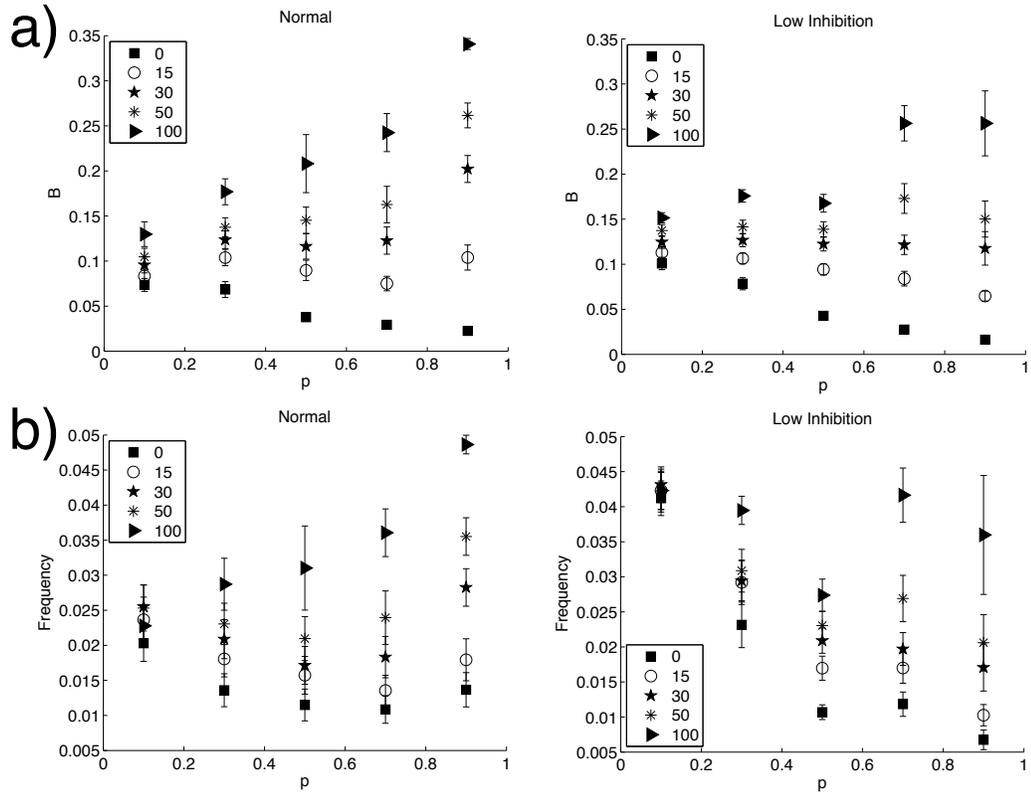}}
\caption{Network dynamics as a function of surviving new neurons and rewiring probability. a) Burst synchronicity measure $B$ as a function of $p$ for several values of the number of incorporated neurons (different markers).  b) Mean firing frequency of networks plotted similarly. In both cases, networks with low rewiring incorporate new cells without significant dynamical changes. Networks with high rewiring show large increases in both frequency and synchronous bursting when new cells are incorporated into them.}
\label{robustPlot}
\end{center}
\end{figure}

It must be noted that our computational model has relatively few biological details. We have far fewer cells than the real hippocampus, no specific cell types, and use a very simplified dynamical model of activity. However, even our simple framework recapitulates many of the important general features of models with considerable biological detail. In healthy conditions, new cells incorporate in functional units that fire given specific stimulus, as in the detailed learning model of \cite{Aimone:2009p7085}. By looking at the frequency of our networks with no new cells in Figure~\ref{robustPlot}, we see a trend toward somewhat higher base activity for more local networks, made much higher by also decreasing inhibition. Similar behavior was reported in extremely detailed and biologically driven hippocampal model of \cite{DyhrfjeldJohnsen:2007p6934}.

Our primary result is a general one that bridges the two behaviors observed during adult neurogenesis. A neurogenic process that is sufficiently activity-dependent to generate stable functional clusters in normal conditions can drive networks with existing pathological connectivity and activity patterns into an \emph{even more pathological} state. Critically, the differences in outcome can be entirely due to differences in the initial neurogenic environment, without any changes in the underlying biochemistry of the neurogenic process. Small-world networks with sufficiently high inhibition were found to be the most robust to activity dependent neurogenesis. Changes to our initial model network that mirrored structural changes to the DG following epileptogenic injury resulted in changes to neurogenic patterns consistent with experimental observation. Connections became less focused and many more cells were able to enter the network. Global synchronous activity also increased. These results suggest that neurogenesis in pathological environments can result in either the development or progression of epilepsy.
		
\bibliographystyle{jneurosci}
\bibliography{ngbiblio}		

\begin{table}\footnotesize
\label{parameterTable}
\caption{Model parameters}
\begin{tabular}{|c|p{4in}|c|}
\hline
Variable & Description & Value taken\tabularnewline
\hline
\hline 
$N_{ex}$ & Number of excitatory neurons & 1000\tabularnewline
\hline 
$N_{in}$ & Number of inhibitory neurons & 200, 100\tabularnewline
\hline 
$N_{add}$ & Number of introduced neurons & 500\tabularnewline
\hline 
$L$ & Number of lattice sites per side & 40\tabularnewline
\hline 
$p$ & Rewiring probability & 0--1\tabularnewline
\hline 
$k_{ee}$ & Average number of excitatory-to-excitatory connections per cell & 35\tabularnewline
\hline 
$k_{ei}$ & Average number of excitatory-to-inhibitory connections per cell & 4\tabularnewline
\hline 
$k_{ie}$ & Average number of inhibitory to excitatory connections per cell & 110\tabularnewline
\hline 
$k_{ii}$ & Average number of inhibitory to inhibitory connections per cell & 4\tabularnewline
\hline 
$m$ & Radial multiplier & 2\tabularnewline
\hline 
$I_{ex}$ & Excitatory cell excitability & 0.73\tabularnewline
\hline 
$I_{in}$ & Inhibitory cell excitability & 0.7\tabularnewline
\hline 
$w_{ee}$ & Excitatory-excitatory connection weight & 0.2\tabularnewline
\hline 
$w_{ei}$ & Excitatory-inhibitory connection weight & 0.4\tabularnewline
\hline 
$w_{ie}$ & Excitatory-excitatory connection weight & -0.4\tabularnewline
\hline 
$w_{ii}$ & Inhibitory-inhibitory connection weight & -0.7\tabularnewline
\hline 
$\alpha$ & Membrane leak constant & 1-1.3\tabularnewline
\hline 
$E$ & External stimulus & 0.4\tabularnewline
\hline 
$\tau_{s}$ & Slow time for spike & 3 ms\tabularnewline
\hline 
$\tau_f$ & Fast time scale for spike & 0.3ms\tabularnewline
\hline 
$\tau_d$ & Spike delay & 0.08 ms\tabularnewline
\hline 
 & Random firing probability per ms & 0.0003\tabularnewline
\hline 
 & Refractory time & 8 ms\tabularnewline
\hline 
 & Integration time step & 0.3 ms\tabularnewline
\hline 
 & Number of stimulated cells & 5\tabularnewline
\hline 
& Excitatory-excitatory connection radius & 6.54 sites\tabularnewline
\hline
& New cell ``nearby" radius in units of the excitory-excitory connection radius & 0.5\tabularnewline
\hline 
 & Number of reconnection chances per cell per connection & 10\tabularnewline
\hline 
 & Time to mature & 2000 ms\tabularnewline
\hline 
 & Threshold firing rate for synapse survival & 30 Hz\tabularnewline
\hline 
 & Threshold firing rate for cell survival & 30 Hz\tabularnewline
\hline 
$a$ & New cell input jitter control parameter & 0.8\tabularnewline
\hline 
 & Time between introduced cells & 350 ms\tabularnewline
\hline 
 & Maximum time window between denoting firing events as causal & 10 ms\tabularnewline
\hline 
\end{tabular}
\end{table}

\end{document}